\documentstyle[aps,prl,multicol,graphicx]{revtex}

\begin{document}
\input{epsf}
\draft
\tighten

\preprint{}
\title{
Zero-field spin splitting in InAs-AlSb quantum wells revisited
}
\author{S. Brosig and K. Ensslin}
\address{Solid State Physics Laboratory, ETH Z\"{u}rich, 8093
Z\"{u}rich,  Switzerland\\
}
\author{R. J. Warburton}

\address{Sektion Physik, LMU M\"unchen, 80539  M\"unchen, Germany\\
}
\author{C. Nguyen, B. Brar, M. Thomas, and H. Kroemer}
\address{Department of Electrical and Computer Engineering, University of 
California, Santa Barbara, Ca 93106, USA\\
}
\date{\today}
\maketitle
\begin{abstract}

We present magnetotransport
  experiments on high-quality InAs-AlSb quantum wells that show 
  a perfectly clean single-period Shubnikov-de Haas oscillation down to very low 
  magnetic fields. In contrast to theoretical expectations based on 
  an asymmetry induced zero-field spin splitting, no beating 
  effect is observed.
  The carrier density has
  been changed by the persistent photo conductivity effect as well as 
  via the application of hydrostatic pressure 
  in order to influence the electric field at
  the interface of the electron gas. Still no indication of spin
  splitting at zero magnetic field was observed in spite of highly
  resolved Shubnikov- de Haas oscillations up to filling factors of
  200. This surprising and unexpected result is discussed 
  in view of other recently published data.
\end{abstract}


\begin{multicols} {2}
\narrowtext
While charge transport in two-dimensional electron gases (2DEG) is
fairly well understood, many open experimental and theoretical
questions related to the spin of the electrons remain. Several
proposals have addressed the possibility of spin transistors, the
detection of Berry's phase, or spin filters in 2DEGs. The standard 2DEG
which is embedded in AlGaAs-GaAs heterostructures is most likely not
the optimal candidate for such investigations, since spin effects as
well as spin-orbit interactions are small perturbations compared to
other effects. This has brought InAs-based material systems into focus
where the electrons reside in an InAs well between AlSb or GaSb
barriers. The unique advantage of this material system in this context
is the large $g$-factor up to $\vert g \vert=15$ and the possibility of large spin-orbit
interactions.

Several experiments in different
material systems 
\cite{Stein83,Stormer83,Luo88,Luo90,Das89,Schultz1986,Nitta97,Engels97,Heida98,Lu98} 
have revealed a 
beating of low-field Shubnikov-de Haas (SdH) oscillations. 
In the literature, these observations have been interpreted as
manifestations of spin-orbit interactions in 
asymmetric quantum wells \cite{Rashba60}. Especially InAs-based systems 
\cite{Luo88,Luo90,Heida98}
are expected to lead to large spin orbit interactions. 
However, Heida et al. \cite{Heida98} found several inconsistencies with 
theoretical expectations. The size of the spin splitting was different for 
samples
from different parts of the wafer, and the spin splitting did not depend on 
the electric field as tuned by a front gate voltage.

In the present paper, we follow up on this question and report
additional inconsistencies, even stronger than those found by 
\cite{Heida98}. We
have conducted SdH studies on many InAs-AlSb quantum wells 
grown by molecular beam epitaxy.
In this paper we focus on samples from four different wafers,
grown by three different individuals, at different times over a 5-year
period, with different $known$ asymmetries. We find the following: 
(a) Tested in the dark, none of the samples shows any SdH beats. 
(b) Under some conditions, beats can be introduced by illumination 
(persistent photoconductivity). (c) The beats under (b) are
strongly sample-size-dependent; they appear only in fairly large
samples, suggesting an essential role of spatial non-uniformities.
With regard to (a), earlier magnetoresistance data by Hopkins et al. 
\cite{Hopkins91} on samples similar to  ours also did not show any SdH 
beats. However, at the time, no particular note was taken of this 
absence, and the matter was not pursued.


All samples contained 15nm-wide InAs quantum wells, 
confined by $AlSb$ or $Al_{x}Ga_{1-x}Sb$ (x $\le$ 0.8) barriers. 
The sample details are summarized in Table 1.
The shutter sequence was designed to enforce InSb-like interfaces \cite{Tuttle90}. 
All growths were on semi-insulating GaAs substrates. To accommodate the $7\%$ 
lattice mismatch between InAs and GaAs
thick ($\ge 1 \mu m$) GaSb buffer layers were grown, 
including a GaSb/AlSb superlattice "smoothing" section \cite{Tuttle90}.
All growths were terminated in a thin (typically $5nm$) cap layer of 
either GaSb or InAs. The nature of the cap, and its (intentional) 
separation from the well via additional electrically inactive spacer layers, 
play an essential role in determining the electron sheet concentreation of the well. 
It is known that the GaSb surface (but apparently not InAs) contains a very 
high concentration of donor-like surface states, at an energy sufficiently high 
to drain electrons into the well \cite{Nguyen93}. For samples grown under otherwise 
identical conditions, the resulting transferred electron concentration 
decreases with increasing well-to-surface distance. In samples 2 and 4, 
these surface states are the dominant source of electrons; neither sample 
contained any intentional doping. In sample 1, with a much deeper well, 
this contribution is small; here the dominant electron source is a 
Te delta-doping donor sheet embedded into the top AlSb barrier;
this is the only sample with intentionally added donors. 
Sample 3 has an InAs cap; 
the electrons in this case are believed to be contributed by donor-like 
interface states at one or both of the well interfaces, or 
interface-related bulk defects in the AlSb barrier; their concentration 
is in good agreement with the values reported by Nguyen et al. \cite{Nguyen93}. 
It is not known how this interface doping is distributed over both interfaces, 
but it is unlikely that the distribution is a symmetrical one. 

The samples were patterned into Hall geometries of $100\,\mu$m width
by wet chemical etching. Voltage probes are placed at several locations along the 
current path, to probe different regions along the sample 
length. Ohmic contacts to the 2DEGs were obtained by
alloying AuGe/Ni contacts. 

\begin{figure}
  \includegraphics[width=8cm]{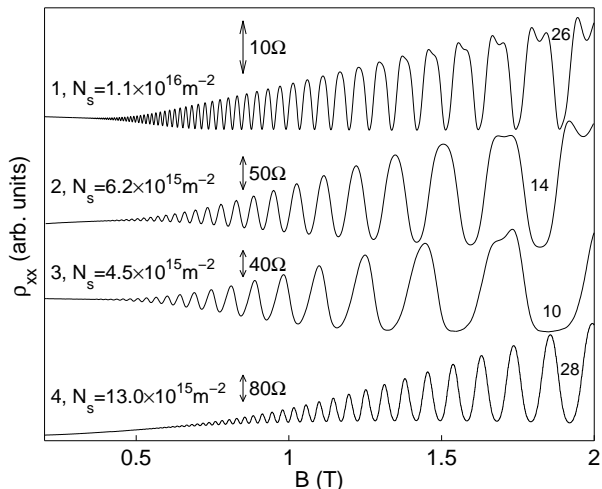}
\caption{Magnetoresistance $\rho_{xx}$ for four different samples taken at
 a temperature of $T=1.7 K$. Shubnikov-de Haas oscillations can be resolved 
 down to magnetic fields of 0.15 T and filling factors up to 200. The 
 numbers 26, 14, 10 and 28 at the right hand side of the figure 
 indicate the postions of the respective filling factors.}
\end{figure}

A magnetic field was applied perpendicular
to the sample surface.  The magnetoresistance of the 4 samples at 
1.7\,K is displayed in Fig. 1.
We have measured the samples at temperatures down to
100\,mK and found no significant improvement of the SdH oscillations,
in agreement with expectations based on estimates of the Landau level
broadening.  Oscillations can be resolved down to magnetic fields of
0.15\,T and filling factors up to 200. All observed features
can be analyzed with one single SdH period with very high accuracy.
From the largest filling factors that we can observe we estimate the
Landau level width to about 0.4\,meV.

An expected zero-field spin splitting should depend on the effective
electric field across the quantum well. Since we found it difficult to
fabricate reliably functioning gates, we varied the carrier
density and with it the effective electric field in the 2DEG via the
persistent photoconductivity effect \cite{Gauer93}. We used a red LED to
illuminate the sample. 
Since we estimate the effective electric field
to be largest in samples 1 and 4, we focus the following discussion on these
samples. Figure 2 displays magnetoresistance traces obtained on sample
1 for three different carrier densities tuned via illumination with light.
The data was taken after the
light was switched off and the carrier density was stable as a
function of time. The Drude scattering time $\tau_{D}$ as obtained 
from the resistivity at B=0 as well as the quantum scattering time 
$\tau_{q}$ from the magnetic field dependence of the SdH amplitude 
are also given for each resistance trace.

The electron density in InAs quantum wells can also be changed by 
hydrostatic pressure \cite{Brosig98}. We reduced the carrier density 
in sample No. 4 by almost a factor of two via application of pressure 
up to $p = 1 GPa$ and did again not find any 
beating pattern in the low-field SdH oscillations (not shown).

\begin{figure}
  \includegraphics[width=8cm]{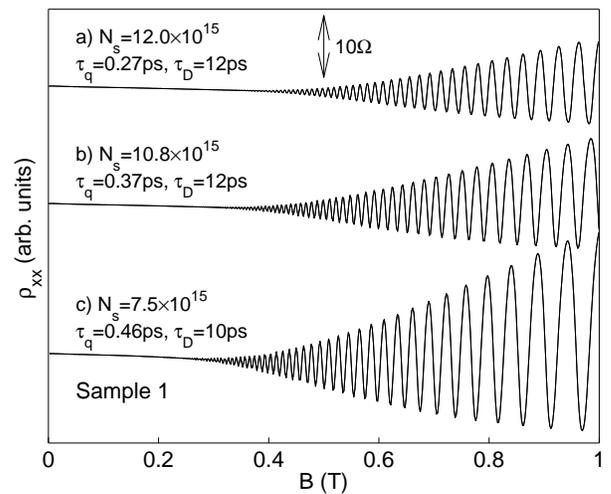}
  \caption{Magnetoresistance $\rho_{xx}$ for sample 1 for three
    different carrier densities changed by the persistent
    photoconductivity effect. The data is taken some time after the
    illumination process such that the carrier density changes by
    less than $10^{-4}$ during the magnetic field sweep. }
\end{figure}

However, in some 
samples, in which
the carrier density could be tuned with light, we found a beating pattern 
right after the illumination. Usually, 
after waiting for some time of the order of an hour the beating pattern was gone.
In a few cases the beating pattern remained constant on the time scales of 
the experiment.
Figure 3 shows resistance traces for sample 1 after 
the sample has been illuminated with an infrared LED and then kept in the 
dark for more than 24 hours. In this stage the resistivity of the sample 
changed by less than $10^{-3}$ per hour. The magnetoresistance across two 
voltage probes separated by 1\,mm clearly displays a weak beating pattern. 
A measurement taken at the same sample at the same time for voltage probes 
separated by only 200\,$\mu$m shows a perfectly one-period SdH pattern. 
Upon further illumination the beating pattern disappeared. We can observe 
such effects very rarely and only for special voltage contacts and 
illumination doses.


There seems to be at least qualitative agreement between experiment and
theory on InAs wells with GaSb barriers \cite{Luo88,Luo90} and other
material systems \cite{Das89,Nitta97,Engels97}. Our data obtained on
InAs quantum wells with AlGaSb barriers with a large Al content 
as well as the data by Hopkins et al. \cite{Hopkins91}
indicating the absence of SdH beating within the
experimental resolution cannot be explained within this framework. The
magnitude of the spin splitting according to the theory of Rashba et
al. \cite{Rashba60} should depend on the effective electric field
across the quantum well. In the following we estimate this
value of the effective electric field for our quantum wells.

\begin{figure}
  \includegraphics[width=8cm]{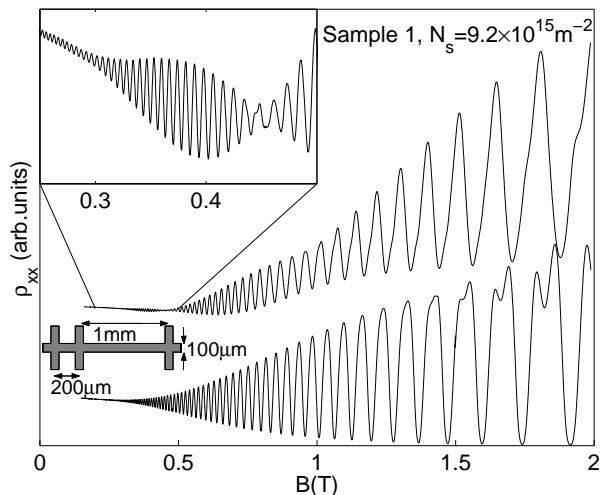}
\caption{Magnetoresistance traces from sample 1 after illumination
    and sufficient stabilization time. A beating pattern is evident in
    the top trace, corresponding to the pair of voltage probes 1\,mm
    apart. There are roughly 21 oscillations between 2 nodes. The
    lower trace, corresponding to the voltage probe pair only
    200\,$\mu$m apart, exhibits no such features, indicating that the
    density in that part of the chip is homogeneous.}
    \end{figure} 

Both the surface states and any Te doping of the top barrier will
introduce a strong transverse electric field into the wells, 
pointing towards the substrate side. If there were no other 
doping sources present, the field at the top of the well would 
be given by $eN_s/\epsilon$, where $N_s$ is the electron sheet concentration, 
and $\epsilon$ the InAs permittivity. The field would decay to zero at the bottom, 
interface, implying an average field of approximately 
$E = eN_s/2\epsilon$.
The background bulk doping in the InAs itself is negligible 
compared to the measured concentrations.
However, part of the electron concentration in all samples is due to 
interface donors, and in sample 3 this is the only known source. 
If we assume that this contribution is symmetrical and has the same 
value in all samples, $4.5 \cdot 10^{15} \, m^{-2}$, we must subtract 
this value from the measured $N_s$. The fields obtained in this way 
are given in the last row of Table I.
If the interface donors were unsymmetrically distributed, 
the values in the Table would have to be adjusted by an amount 
depending on the magnitude and sign of the asymmetry, 
maximally $\pm 3.5 \cdot 10^6 V/m$, but probably much less.

With the possible exception of sample 3, all samples have large built-in
asymmetries, with transverse electric fields estimated to range from
$6.6 \cdot 10^6 \, V/m$ to $5.0 \cdot 10^6 \, V/m$ for Samples 4 and 1, 
down to nominally zero for sample 3. The uncertainties on these 
estimates are on the order $\pm 1 \cdot 10^6 \, V/m$, i.e. small 
compared to the range of values. It is extraordinarily unlikely 
that accidental effects would compensate the different asymmetries 
in all samples. The absence of the SdH beating in our samples as
well as those of Hopkins et al. \cite{Hopkins91} suggests a more 
fundamental suppression mechanism, somehow associated with InAs/AlSb wells, 
but absent in GaAs/(Al,Ga)As wells, and even in InAs/GaSb wells. 
The data of Heida et al. \cite{Heida98} appear to contradict this 
hypothesis, but it may be important that even their work indicates 
significant discrepancies between experiment and theory.

A Fourier-transform of the SdH pattern of sample 1 indicates a resolution 
of our experiment of better than 1\,meV for the possible detection of a 
beating phenomenon. This limit is comparable with the one obtained from
the width of the Landau levels.
We have self-consistently calculated \cite{Snider} the conduction 
band profile and wave function based on the sample parameters 
and then calculated the expected spin splitting using Rashba`s theory \cite{Rashba60}.
We found a value of about 5 \, meV in agreement with Refs.
\cite{Lommer85,Lommer88,Silva94}.

Let us now return to the light induced beating pattern as displayed in Fig. 3.
As light changes the 
carrier density, it also changes the effective electric field across the 
well. If this were the underlying reason for the observed beating pattern 
one would expect that the beating pattern is present without light, 
disappears at some does of light as the potential well becomes symmetric and 
then appears again once the asymmetry points to the other direction. In our 
case, if we 
observe this feature at all in an experiment, the beating pattern is only 
present for a certain dose of light, it is absent for lower and higher 
carrier densities. These observations strongly hint at the fact that in our 
samples a beating pattern in the low-field SdH oscillations does not stem 
from an asymmetry induced Rashba-type interaction.
 
In the following we argue that the observed SdH beating pattern in 
Fig. 3 arises from 
an inhomogeneous carrier 
distribution induced by the illumination.
The light is not distributed homogenously along 
the Hall geometry and might therefore lead to an inhomogeneous carrier 
distribution. If a reasonable number of areas of different carrier density 
occur along the 
current path of the Hall geometry this could lead to a beating pattern of 
the low-field SdH oscillations. After the carriers have had enough time 
to relax back to thermal 
equilibrium the inhomogeneities and with it the beating pattern disappear.
The time scales of the non-persistent photoconductivity effect are of the 
order of hours and are consistent with the disappearance of the 
beating pattern.

The importance of sample inhomogeneities obviously depends on the 
length scale of the 
experiment. The data in Fig. 3 suggest that over short length scales, in 
this case 200\,$\mu$m, the sample is homogeneous within the experimental 
resolution and therefore displays single period SdH oscillations. 
For a larger length scale of 1\,mm the 
beating pattern is experimentally observed. We find roughly 21 oscillations 
between two nodes of the beating. If interpreted in terms of 
sample inhomogeneities this leads to a value of 
$\Delta N_s / N_s \approx 5\%$, 
which is not an unreasonable number.

While we do not question the valid interpretion of other experiments in 
terms of the Rasha-type spin orbit splitting, our experimental results 
cannot be explained within this framework. It is not clear why in our 
InAs-AlSb quantum wells the low-field SdH beating cannot be observed.

We do not know why our samples behave
differently compared to Ref. \cite{Heida98} but like to stress that 
our sample quality is higher in
terms of scattering times and electron mobilities. We do not
expect to observe Berry phase-type effects in our samples
\cite{Morpungo98} induced by strong Rashba-type spin orbit
interaction.

From Fig. 1 it is obvious that spin splitting of SdH oscillations can be 
observed at magnetic fields as low as $B=1.5 T$. The magnitude of the 
g-factor in our quantum wells can be determined by temperature dependent 
measurements or via experiments where the magnetic field in tilted with 
respect to the sample surface. We find in both cases values for the g-factor 
of $ \vert g \vert \approx 12-15$ \cite{Brosig98a}. This makes InAs-AlSb quantum wells 
promising candidates for spin-related experiments. 

The fact that we do not observe a beating of the low-field SdH oscillations 
comes as a surprise 
and is completely unexpected. While spin-orbit 
interaction in general could still play a substantial role in these systems 
the contribution of the quantum well inversion asymmetry to it is likely to 
be small. This, however, could be an advantage for the possible realization 
of coupled spin states in quantum dots. \cite{Loss98}

We are grateful to T. Heinzel and S. Ulloa for helpful discussions and
thank ETH Z\"urich and QUEST for financial support.

\vspace{-0.6cm}
\begin{table}
\caption{Summary of parameters describing the layer sequence and the 
  electronic properties of the samples at T=1.7 K}

\def\boxit#1{\parbox[t]{3cm}{\baselineskip=10pt
\hangafter 1\hangindent8pt{#1}}}

\begin{tabular}{ldddd}
Sample&  No. 1       & No. 2          & No. 3       & No. 4    \\
\hline
\boxit{UCSB ID}         
        & 9110-52         & 9503-18             & 9401-38          & 9602-24      \\
\boxit{Distance of InAs well to surface (nm)}         
        & 215         & 28             & 28          & 56      \\
Cap material
        & GaSb        &GaSb            & InAs        & GaSb    \\
\boxit{Electron Density \\ ($10^{15} $m$^{-2}$)}
       & 11.0         & 6.2            & 4.5         & 13.0    \\
\boxit{Electron mobility\\ (m$^2$/Vs)}  
       &   70         &    84          &  28         & 42      \\
\boxit{Drude scattering\\ time (ps)}   
       & 12           &     14         &   4.8       & 7       \\
\boxit{Quantum  scattering\\
time (ps)} 
       & 0.27         &     0.18       &    0.16     &  0.14   \\
\boxit{Estimated el. field \\ ($10^{6}$V/m)}  
       & 5.0           & 1.3            & 0.0         &  6.6     \\
\end{tabular}
\end{table}

\end{multicols}

\end{document}